\documentclass[10 pt, conference]{ieeeconf}  

\IEEEoverridecommandlockouts                              
\usepackage{graphics} 
\usepackage{epsfig} 
\usepackage{mathptmx} 
\usepackage{times} 
\usepackage{amsmath} 
\usepackage{amssymb}  
\usepackage{subfig}

\usepackage{cite}

\usepackage{stackengine}

\title{\LARGE \bf
Analysis of Fairness-promoting Optimization Schemes of Photovoltaic Curtailments for Voltage Regulation in Power Distribution Networks 
}
\author{Rahul K. Gupta and Daniel K. Molzahn
\thanks{School of Electrical and Computer Engineering, Georgia Institute of Technology, Atlanta, USA, 30308. Email: \{rahul.gupta, molzahn\}@gatech.edu. D.K. Molzahn acknowledges support from the NSF AI Institute for Advances in Optimization (AI4OPT), \#2112533.}
}
\begin{document}
\newcommand\barbelow[1]{\stackunder[1.2pt]{$#1$}{\rule{.8ex}{.075ex}}}
\maketitle
\thispagestyle{empty}
\pagestyle{empty}
\begin{abstract}
Active power curtailment of photovoltaic (PV) generation is commonly exercised to mitigate over-voltage issues in power distribution networks. However, fairness concerns arise as certain PV plants may experience more significant curtailments than others depending on their locations within the network. Existing literature tackles this issue through fairness-promoting/aware optimization schemes. These schemes can be broadly categorized into two types. The first type maximizes an additional fairness objective along with the main objective of curtailment minimization. The second type is formulated as a feedback controller, where fairness is accounted for by assigning different weights (as feedback) in the curtailment minimization objective for each PV plant based on previous curtailment actions. In this work, we combine these two schemes and provide extensive analyses and comparisons of these two fairness schemes. We compare the performance in terms of fairness and net curtailments for several benchmark test networks.
\end{abstract}
\section{INTRODUCTION}
The accelerated penetration of photovoltaic (PV) generation in power distribution networks is causing several problems, especially concerning the operation of the grid under the network's physical constraints \cite{guide2004voltage, CIGREREF, IEEE_practice}. The existing literature (e.g., \cite{seuss2015improving, ding2016technologies, hashemi2016efficient, tavares2020innovative}) addresses this challenge by intelligently controlling these PV plants so that grid constraints are always respected. These control schemes include both active and reactive power regulation from the PV inverters to address over-voltage issues, as seen in \cite{seuss2015improving, home2022increasing, luthander2016self, von2018strategic, sevilla2018techno, o2020too}. These schemes aim to minimize curtailment for every power plant; however, the inherent physics of the power distribution system leads to disparities in curtailments based on the locations of the connected PV units in the network. For instance, customers located at the end of the feeder face more curtailments compared to those near the substation. This problem has been highlighted in \cite{ali2015fair}.

Recent works have increasingly studied fairness in the context of PV curtailments and proposed different fairness-promoting/aware control schemes \cite{ali2015fair, Lusis2019Reducing, liu2020fairness, gebbran2021fair, gerdroodbari2021decentralized, wei2023model, gupta2024fairness}. These methods differ in how they enforce fairness in PV control algorithms. The work in \cite{liu2020fairness} evaluates different objectives (maximize self-consumption, energy exported, and financial benefit) in terms of achieved fairness. The work in \cite{ali2015fair} proposes fair power curtailment by exploiting sensitivity matrix information in a P-V droop control scheme. In \cite{Lusis2019Reducing}, an additional cost term is added to the curtailment minimization problem; this term reduces the variance of the curtailments across different PV plants. In \cite{gebbran2021fair}, a fairness cost function is introduced, aiming to curtail proportionally to the energy exported. 
In \cite{wei2023model}, a model-free control scheme is proposed, where fairness is accounted for by different objectives, one of which is fairness in curtailed PV proportionally to maximum available generation. In \cite{gupta2024fairness}, extra objectives are utilized to minimize disparity in curtailments among different PV owners using day-ahead forecasts of PV generation and demand.

To summarize, most of the existing literature on fairness-promoting schemes can be categorized into two types. The first is by incorporating a fairness cost function into the optimization problem (e.g., \cite{Lusis2019Reducing, gebbran2021fair, poudel2023fairness, gupta2024fairness}) 
These schemes effectively enhance fairness in curtailments; however, they necessitate proper tuning of the weight parameter assigned to the fairness objectives. The solution of such control schemes heavily depends on the weights assigned to the fairness objective, and the outcome could be very sensitive to these weights as illustrated in \cite{gupta2024fairness}. These schemes do not account for the realization of past curtailments in the fairness formulation.
The second refers to the authors' previous work in \cite{gupta2024improving} that proposed a \emph{feedback-based approach}, where the curtailment actions in previous time steps are used to determine the fairness-informed weights in the curtailment minimization problem, penalizing the PV plants that were not curtailed previously. Such a feedback scheme allows for increased fairness over time by considering previous curtailment actions, which contrasts with other schemes that do not account for past curtailments. However, an extensive comparison between the two types of fairness schemes has not yet been studied.

\begin{figure}[!t]
    \centering
    \includegraphics[width=\linewidth]{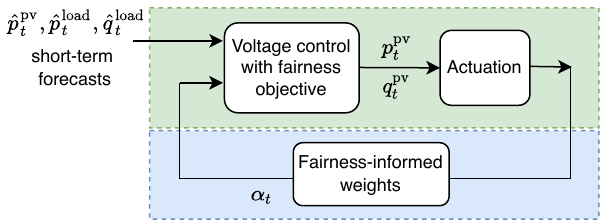}
    \caption{Flow diagram of the different fairness-promoting PV curtailment schemes. Green refers to those with additional fairness objective and green \& blue refers to a feedback-based approach along with additional fairness objectives.}
    \vspace{-1em}
    \label{fig:flowv2}
\end{figure}

In this context, this paper aims to provide extensive analyses and comparisons of the two fairness-aware formulations. We combine the two formulations into a single scheme containing both the additional fairness objective and as well as feedback-based individual weights using previous curtailment decisions. Such a scheme allows us to compare the fairness versus curtailment outcomes for different choices of weight factors. We do this via a Pareto curve computed by sweeping across the range of weight factors that balance PV curtailments versus fairness. The combined scheme is illustrated schematically in Fig.~\ref{fig:flowv2}. The green refers to the voltage control problem with an extra fairness objective, while the blue refers to the case when the fairness is implemented by some kind of feedback of the previous realizations.

The key contributions of this work are listed below.
\begin{enumerate}
    \item A fairness-aware optimization scheme that integrates the additional fairness objectives as well as feedback-based approach for determining fairness-informed weights using the previous curtailment actions.
    \item Numerical simulation of the feedback-based approach with different objectives such as minimizing the normalized electricity bill (or maximizing the export of PV generation) and simply minimizing the curtailments.
    \item A performance comparison of two fairness-promoting methodologies, both incorporating fairness objectives: one without feedback consideration and the other integrating feedback. Analysis includes different fairness objectives, with one accounting for previous curtailments and the other not.
\end{enumerate}




The paper is organized as follows: Section~\ref{sec:Preliminary} formulates the voltage control problem, Section~\ref{sec:fairness_aware} describes different fairness-promoting/aware formulations and the proposed feedback-based approach, Section~\ref{sec:numerical_validation} presents numerical simulations, and finally, Section~\ref{sec:conclusion} concludes the paper.

\section{PRELIMINARIES}
\label{sec:Preliminary}
We consider a power distribution network with multiple PV generators that are causing over-voltage problems. 
The objective of the voltage control problem is to regulate the active and reactive power injections from the PV plants such that the nodal voltages are respected within the operational limits. In the following, we detail each component of the voltage control problem, i.e., the power grid model, the PV controllability model, and the voltage control problem.
\subsection{Grid Model}
Let the distribution network be composed of $N$ nodes; let the symbol $\mathcal{N}$ denote the set of nodes in the network, i.e., $\mathcal{N} = \{1, \dots, N\}$. The network has multiple uncontrollable loads as well as PV plants. Let the net active and reactive power injections per node $i$ be denoted by $p_{i}$ and $q_{i}$, respectively. Let the symbols $\mathbf{p}$ and $\mathbf{q}$ contain the nodal active and reactive powers for all the nodes in $\mathcal{N}$. The complex nodal voltages and voltage magnitudes are denoted by symbols $v_i$ and $|v_i|$, respectively, for each node $i$.

Using the power-flow equations, the voltage for node $i$ can be represented by a symbolic non-linear function $\mathcal{V}_i$
\begin{align}
    |v_i| = \mathcal{V}_i(\mathbf{p}, \mathbf{q}, v_0) ,\label{eq:non_linear_pf}
\end{align}
where $v_0$ refers to the reference voltage (usually the feeder voltage) in distribution systems.

The original non-linear power-flow model in \eqref{eq:non_linear_pf} leads to a non-linear and non-convex formulation, referred to as the classical optimal power problem (OPF). Often such a formulation is either relaxed or linearized for the sake of tractability. These schemes have their advantages and disadvantages~\cite{molzahn2019survey}. We use a power-flow linearization approach to obtain a linear optimization problem, although any other power-flow model can be used. We rely on the first-order Taylor's approximation where the linearization coefficients are updated based on the grid measurements from the last timestep. Assuming that the operating point for Taylor's linearization is denoted by $|v_i^\bullet|, p_j^\bullet, q_j^\bullet$, the voltage magnitudes are linearized as
\begin{align}
    |v_i| = |v_i^\bullet| + \sum_{i \in \mathcal{N}}K_{ij}^p(p_j - p_j^\bullet) +  \sum_{i \in \mathcal{N}}K_{ij}^q(q_j - q_j^\bullet),\label{eq:vmodel}
\end{align}
where the coefficients $K_{ij}^p, K_{ij}^q$ are the partial derivatives of the voltage magnitudes at node $i$ with respect to the injections at node $j$. They are given as 
\begin{align}
    K_{ij}^p  = \frac{\partial |v_i|}{\partial p_j} ~\text{and}~  K_{ij}^q  = \frac{\partial |v_i|}{\partial q_j} .
\end{align}
These coefficients $K_{ij}^p, K_{ij}^q, \forall i,j\in \mathcal{N}$, are computed and updated for each time-step based on the most recent realizations of the active and reactive power injections. Here, we assume the ability to access to the most recent state using a real-time state estimation process. The sensitivity coefficients are computed using the approach proposed in \cite{christakou2013efficient} that requires information on the nodal voltages and the compound admittance matrix of the network. This scheme allows for uniquely computing these coefficients by solving systems of linear equations presented in \cite{christakou2013efficient}, which provides a unique solution for every operating point as long as the load-flow Jacobian is locally invertible.
\subsection{Photovoltaic Control Model} We assume that the PV plants connected to the electrical network can be regulated with respect to their active and reactive power outputs. Let the number of PV plants be denoted by $N^\text{pv}$ and their bus indices be contained in set $\mathcal{N}^{\text{pv}}$. 
Let $t \in \mathcal{T}$ represent the time index and $\mathcal{T}$ be the set of time indices during a day. Let the active and reactive power from PV plants be denoted by $p_{l,t}^\text{pv}$ and $q_{l,t}^\text{pv}$, respectively. 
The active power injections are constrained by the PV power potential modeled by the maximum power point tracking (MPPT) denoted by $\hat{p}_{l,t}^\text{pv}$. The corresponding constraint is
\begin{align}
     0 \leq p_{l,t}^\text{pv} \leq \hat{p}_{l,t}^\text{pv}. \label{eq:PV_mpp}
\end{align}
The MPPT $\hat{p}^\text{pv}$ is modeled by a short-term forecasting tool, similar to the ones described in \cite{gupta2024fairness}.

The reactive power outputs from the PV inverters are constrained by the power-factor limits given as
\begin{align}
       & q_{l,t}^\text{pv} \leq \xi p_{l,t}^\text{pv} ~~\text{and} ~ 
        -q_{l,t}^\text{pv} \geq \xi p_{l,t}^\text{pv},
       \label{eq:PV_Q}
\end{align}
where $\xi$ imposes the power-factor constraint on the PV inverter, which is a parameter in the problem. For example, $\xi = 0.33$ for a power factor of 0.95.

The active and reactive power outputs are also constrained by the PV inverters; nominal ratings ($S^\text{pv}$) given by 
\begin{align}
    &  (p_{l,t}^\text{pv})^2 +  (q_{l,t}^\text{pv})^2 \leq (S^\text{pv})^2,
\end{align} 
which we piece-wise linearize to obtain a linear set of constraints; they are re-written as:
\begin{align}
    & m_k ({p}_{l, t}^{\text{pv}}) + {q}_{l, t}^{\text{pv}} \leq  n_k, && k = 1, \dots, K, &  \label{eq:capability_linear}
\end{align}
where $m_k$ and $n_k$ denote the linearization coefficients for $k$-th segment in the piece-wise linearization and $K$ denotes the number of linear segments.
\subsection{Demand Model}
\label{sec:demand}
We model the electricity demand to be uncontrollable but predictable based on short-term forecasts. We assume the ability to access measurements of the load that we use as forecasts using a persistent forecasting policy, i.e., we use the last-known measurements as a proxy for the forecast of the load in the next time step. This approach works well for real-time control schemes as demonstrated in previous works~\cite{gupta2020grid, gupta2022reliable}. 
The forecasts for active and reactive loads for node $n \in \mathcal{N}_\text{load}$ are denoted by $\hat{p}^\text{load}_{n, t}$ and $\hat{q}^\text{load}_{n, t}$, respectively, for time $t$.

\subsection{Voltage Control Problem}
\label{sec:unfair_case}
The voltage control problem aims to satisfy the nodal voltage constraints in the power distribution networks by curtailing PV generation and providing reactive power regulation.
The voltage control problem is formulated as the minimization problem:
\begin{subequations}
\label{eq:v_control_prob}
\begin{align}
\begin{aligned}
    \label{eq:obj_func_unfair}
   \underset{p_{l,t}^\text{pv}, q_{l,t}^\text{pv}}{\text{minimize}}~ 
   \sum_{l\in\mathcal{N}^{\text{pv}}} 
   f_l
\end{aligned}
\end{align}
subject to:
\begin{align}
    & v^{min} \leq |v_{i,t}| \leq v^{max}  && \forall i \in \mathcal{N}, t\in\mathcal{T}\\
    & Eq.~\eqref{eq:PV_mpp}, \eqref{eq:PV_Q}, \eqref{eq:capability_linear} && \forall i \in \mathcal{N}, t\in\mathcal{T}.
\end{align}
\end{subequations}
Here, the term $f_l$ 
refers to the objective function which is related to the controllability of the PV units. We will describe two different objectives later in this section. 
\section{FAIRNESS-PROMOTING\\ VOLTAGE CONTROL SCHEMES}
\label{sec:fairness_aware}
As mentioned before, we evaluate two different fairness-aware schemes. We then present a combined formulation.
\subsection{Fairness by An Additional Objective Term}
\label{sec:fairness_w_obj}
This approach adds an extra fairness objective to the objective function of \eqref{eq:obj_func_unfair}, similar to \cite{Lusis2019Reducing, gebbran2021fair, poudel2023fairness, gupta2024fairness}:
\begin{align}
\begin{aligned}
   \sum_{l\in\mathcal{N}^{\text{pv}}} \Big(f_l + w\big\|\gamma - h_l({p}_{l}^{\text{pv}}, \theta_l) 
     \big\|_1\Big).
\end{aligned}
\end{align}
Here, $\gamma$ is a variable that enforces fairness among different PV plants with respect to function $h_l({p}_{l}^{\text{pv}}, \theta_l)$ (described later). The symbol $w$ denotes the weight assigned to the fairness objective. The symbol $\theta_l$ encapsulates all the known quantities such as the realizations from previous timesteps and forecasts.

\subsection{Fairness as A Feedback Controller}
\label{sec:feedback_control}
As proposed in the authors' previous work in \cite{gupta2024improving}, this formulation accounts for fairness by a feedback-based approach where the previous curtailments are used to determine weights to the individual $f_l$ in \eqref{eq:obj_func_unfair}:
\begin{align}
\begin{aligned}
    \label{eq:obj_func}
   \sum_{l\in\mathcal{N}^{\text{pv}}} \alpha_{l,t} 
   f_l
\end{aligned}
\end{align}

Here, the factors $\alpha_{l,t}$ are \emph{fairness-informed weights} that are assigned per PV plant; these weights are decided to enforce fairness in the curtailments decisions for each timestep based on the realization of the curtailments in the preceding timesteps. The intuition behind such weights is that the PV plants that were heavily curtailed in the previous timesteps (cumulatively) will be less preferred to be curtailed in the upcoming timesteps. Such a scheme informs the optimizer about potential unfairness that occurred in the previous operations in order to take necessary actions in the next set-points, as described in the following subsections. 
\subsection{Combining the Two Formulations}
\label{sec:combined_formulation}
We combine the two formulations above into one single scheme. Such a scheme will allow us to evaluate the importance of formulation and perform an extensive comparison. The combined formulation is
\begin{align}
\begin{aligned}
   \sum_{l\in\mathcal{N}^{\text{pv}}} \Big(\alpha_l f_l + w\big\|\gamma - h_l({p}_{l}^{\text{pv}}, \theta_l) 
     \big\|_1\Big).
\end{aligned}
\end{align}

In the following, we describe two different objectives $f_l$ and corresponding $h_l$, $\alpha_l$.
\subsection{Different Objectives and Fairness Functions}
\subsubsection{Minimizing the Net Electricity Bill}
We first consider an objective that minimizes the electricity cost of the consumers or, in other words, maximizes the benefit earned by the PV generators. We assume that the consumers pay a fixed electricity tariff rate $c^\text{im}$ and can receive remuneration with feed-in tariff rate $c^\text{fit}$. Then, the price of electricity per plant at time $t$ is 
\begin{align}
        f^{\text{bill}}_l = c^\text{im}p_{l,t}^\text{im} - c^\text{fit} p_{l,t}^\text{ex}, \label{eq:net_price}
 \end{align}
where $p_{l,t}^\text{im}$ and $p_{l,t}^\text{ex}$ refer to the imported and exported electricity, which are expressed as
\begin{align}
    & p_{l,t}^\text{im} = [\hat{p}^\text{load}_{l,t} - p^\text{pv}_{l,t}]^+~~~ \text{and}
    & p_{l,t}^\text{ex} = [-(\hat{p}^\text{load}_{l,t} - p^\text{pv}_{l,t})]^+
\end{align}
where the operator $[u]^+ = \text{max}(u,0)$.
Thus, the expression in \eqref{eq:net_price} can be rewritten as 
\begin{align}
    f^{\text{bill}}_l = c^\text{im}[\hat{p}^\text{load}_{l,t} - p^\text{pv}_{l,t}]^+ - c^\text{fit}[-(\hat{p}^\text{load}_{l,t} - p^\text{pv}_{l,t})]^+. \label{eq:net_price2}
\end{align}
Furthermore, the expression $[\hat{p}^\text{load}_{l,t} - p^\text{pv}_{l,t}]^+$ can be re-written as $(\hat{p}^\text{load}_{l,t} - p^\text{pv}_{l,t}) + [-(\hat{p}^\text{load}_{l,t} - p^\text{pv}_{l,t})]^+$. Therefore, \eqref{eq:net_price2} can be written as: 
\begin{align}
    f^{\text{bill}}_l = (c^\text{im} - c^\text{fit})[-(\hat{p}^\text{load}_{l,t} - p^\text{pv}_{l,t})]^+ +c^\text{im}(\hat{p}^\text{load}_{l,t} - p^\text{pv}_{l,t}). \label{eq:net_price3}
\end{align}

The objective in \eqref{eq:net_price3} is convex as the sum of two convex functions, provided that $c^\text{im} \geq c^\text{fit}$, which is usually the case in power distribution systems. We denote the objective in \eqref{eq:net_price3} as $f^{\text{bill}}_l({p}^\text{pv}_{l,t}, {p}_{l,t}^\text{load})$. 

In this case, the weights $\alpha_{l,t}$ are decided based on the net earnings that the consumers can make by exporting PV generation. We define a metric as the ratio of the earnings due to PV export for each consumer (with curtailment) vs. the total earnings that could have been made without any curtailments. This metric is defined as:
\begin{align}
\label{eq:fairness_bill_func}
\mathcal{E}_l =
    \frac{\sum_{\tau=1}^{t-1} \big(-f^{\text{bill}}_l(\tilde{p}^\text{pv}_{l,\tau}, \tilde{p}_{l,\tau}^\text{load})\big)}{\sum_{\tau=1}^{t-1} \big(-f^{\text{bill}}_l(\tilde{\hat{p}}^\text{pv}_{l,\tau}, \tilde{p}_{l,\tau}^\text{load})\big)},
\end{align}
where $\tilde{p}^\text{pv}_{l,t}$ and $\tilde{p}^\text{load}_{l,t}$ refer to realized PV and load active powers during actual operations, which might be different from the setpoint ${p}^\text{pv}_{l,t}$ and forecasts due to forecast uncertainties. The symbol $\tilde{\hat{p}}^\text{pv}_{l,t}$ denotes the estimated PV MPP that is computed by observed global horizontal irradiance using PV model described in \cite{sossan2019solar}. The weight $\alpha_{l,t}$ is defined as the inverse of the normalized earnings $\mathcal{E}_l$ to favor the nodes where PV units are heavily curtailed, resulting in reduced earnings. This weight is
\begin{align}
    \alpha^{\text{bill}}_{l,t}  = {1}/{\mathcal{E}_{l,t}}.
\end{align}

The fairness function $h_l$ is defined as the ratio of the optimized electricity bill with and without curtailments, and can be defined in two ways: 
\paragraph{Not considering past curtailments}
\begin{align}
\label{eq:objective_w_fairnessp0}
h_l^{\text{bill}}({p}_{l}^{\text{pv}}, \theta_l)  = 
\frac{f^{\text{bill}}_l({p}^\text{pv}_{l,t},\hat{p}_{l,t}^\text{load})}{ f^{\text{bill}}_l(\hat{p}^\text{pv}_{l,t},\hat{p}_{l,t}^\text{load})}.
\end{align}
\paragraph{Considering past curtailments}
\begin{align}
\label{eq:objective_w_fairness}
h_l^{\text{bill}}({p}_{l}^{\text{pv}}, \theta_l)  = 
    \frac{\sum_{\tau=1}^{t-1} \big(f^{\text{bill}}_l(\tilde{p}^\text{pv}_{l,\tau}, \tilde{p}_{l,\tau}^\text{load})\big) + f^{\text{bill}}_l({p}^\text{pv}_{l,t},\hat{p}_{l,t}^\text{load})}{\sum_{\tau=1}^{t-1} \big(f^{\text{bill}}_l(\tilde{\hat{p}}^\text{pv}_{l,\tau}, \tilde{p}_{l,\tau}^\text{load})\big) + f^{\text{bill}}_l(\hat{p}^\text{pv}_{l,t},\hat{p}_{l,t}^\text{load})}.
\end{align}
\subsubsection{Minimize PV Curtailments}
The objective in this case is to minimize the net curtailments with respect to the PV MPPT potential. This objective is formulated as the difference between the potential generation and the optimized one as:
\begin{align}
    f^{\text{curt}}_l = \hat{p}_{l,t}^\text{pv} - p_{l,t}^\text{pv}.
\end{align}

In this case, we define a metric to quantify the PV curtailments as the ratio of the realized PV generation with respect to the PV MPP potential; that is:
\begin{align}
\label{eq:fairness_curt_func}
    \mathcal{G}_{l,t} = \frac{\sum_{\tau=1}^{t-1}\tilde{p}_{l,\tau}^{\text{pv}}}{\sum_{\tau=1}^{t-1}\tilde{\hat{p}}_{l,\tau}^{\text{pv}}}.
\end{align}
In this case, the weight $\alpha_{l,t}$ is defined as 
\begin{align}
    \alpha_{l,t} = {1}/{\mathcal{G}_{l,t}}
\end{align}
which penalizes the PV plants that were not curtailed during the previous timesteps. 

In this case, $h_l$ is defined as the ratio of generation with and without curtailment. It can be also defined in two ways:
\paragraph{Not considering past curtailments}
\begin{align}
\label{eq:objective_w_fairness_curtp0}
   h_l^{\text{curt}}({p}_{l}^{\text{pv}}, \theta_l)  =  \frac{{p}_{l,t}^{\text{pv}}}{ \hat{p}_{l,t}^{\text{pv}}}.
\end{align}
\paragraph{Considering past curtailments}
\begin{align}
\label{eq:objective_w_fairness_curtp1}
   h_l^{\text{curt}}({p}_{l}^{\text{pv}}, \theta_l)  =  \frac{\sum_{\tau=1}^{t-1}\big(\tilde{p}_{l,\tau}^{\text{pv}}\big) + {p}_{l,t}^{\text{pv}}}{\sum_{\tau=1}^{t-1} \big(\tilde{\hat{p}}_{l,\tau}^{\text{pv}}\big) + \hat{p}_{l,t}^{\text{pv}}}.
\end{align}

In the following sections, we will present performance comparisons of the above schemes by selecting different fairness functions and varying weights associated with them.
\section{NUMERICAL SIMULATIONS}
\label{sec:numerical_validation}
We next present numerical simulations of different fairness-promoting schemes on multiple benchmark networks. We first show results for the feedback-based approach (i.e., Sec.~\ref{sec:feedback_control}) and compare against the case that does not consider fairness (Sec.~\ref{sec:unfair_case}) for different benchmark networks. Then, we compare the performance of the two formulations with additional objectives accounting and not accounting for the fairness by feedback-based approach. We also present a performance comparison of different fairness objectives: one that accounts for past curtailment decisions and another that does not.
\subsection{Benchmark Testcases}
\subsubsection{CIGRE Low Voltage (LV) Benchmark Setup}
\begin{figure}[!htbp]
    \centering
\includegraphics[width=\linewidth]{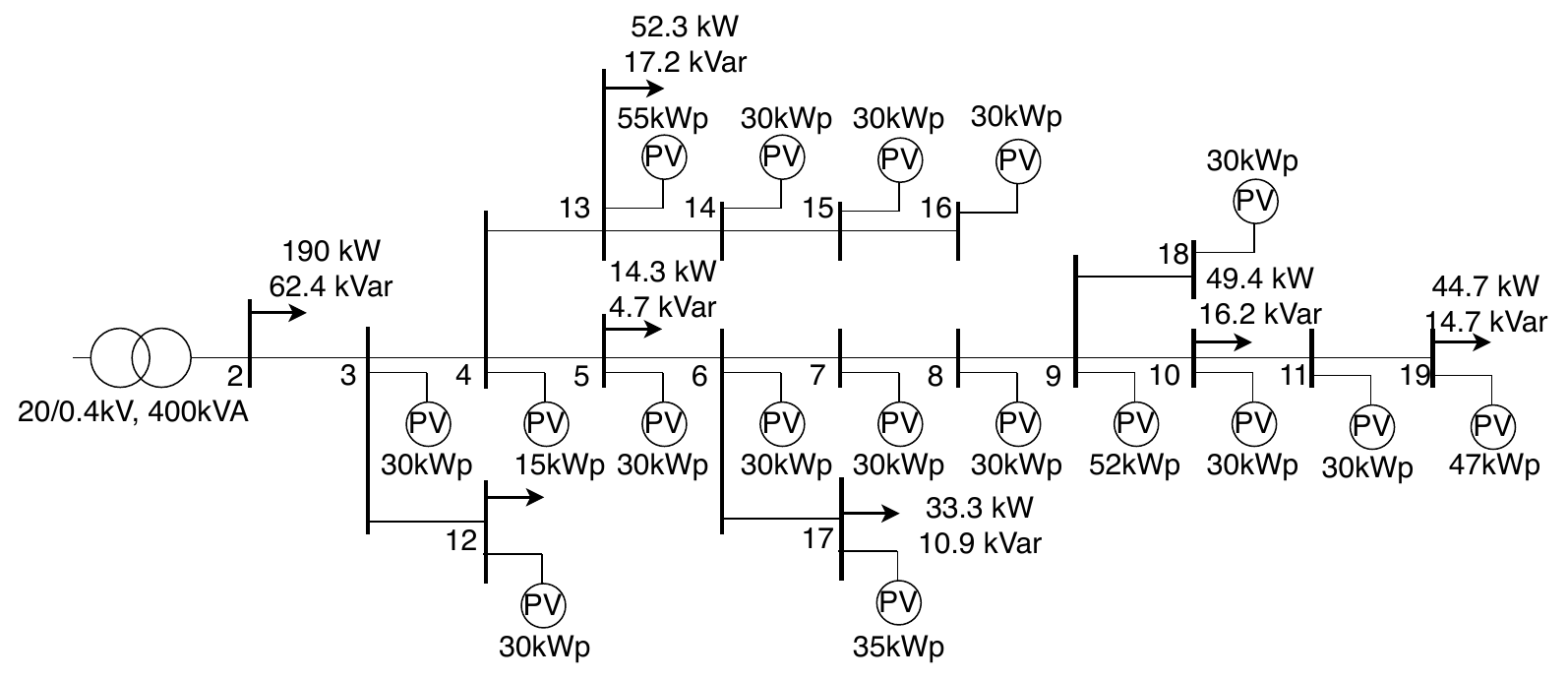}
    \caption{CIGRE low voltage benchmark network installed with multiple PV plants.}
    \label{fig:cigrelv}
\end{figure}

Fig.~\ref{fig:cigrelv} shows the CIGRE benchmark \cite{CIGREREF} network which is a low-voltage system connected to the upstream system via a 20/0.4kV, 400kVA transformer. This figure also shows the nominal active and reactive load demands per node. We connect several PV plants to create over-voltage issues in the network so that we can show the effectiveness of the proposed scheme with significant PV curtailments. The capacities of these PV plants are also shown in Fig.~\ref{fig:cigrelv}.
\begin{figure}[!htbp]
\centering
\subfloat[PV Generation]{\includegraphics[width=\linewidth]{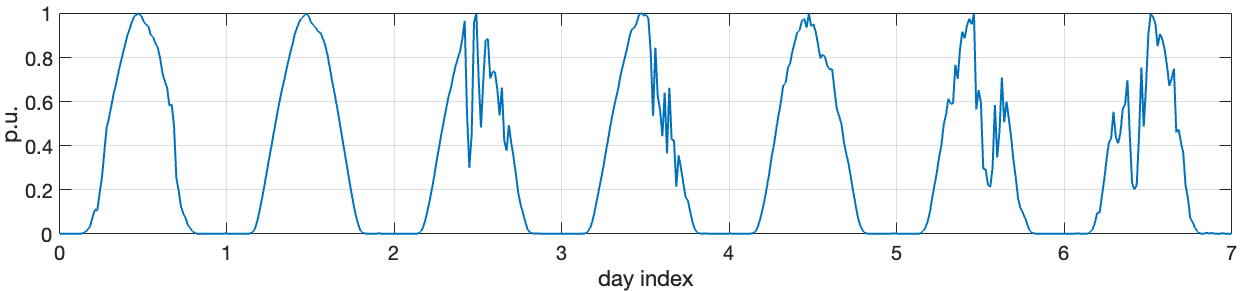}
\label{fig:PV}}\\
\subfloat[Load Demand]{\includegraphics[width=\linewidth]{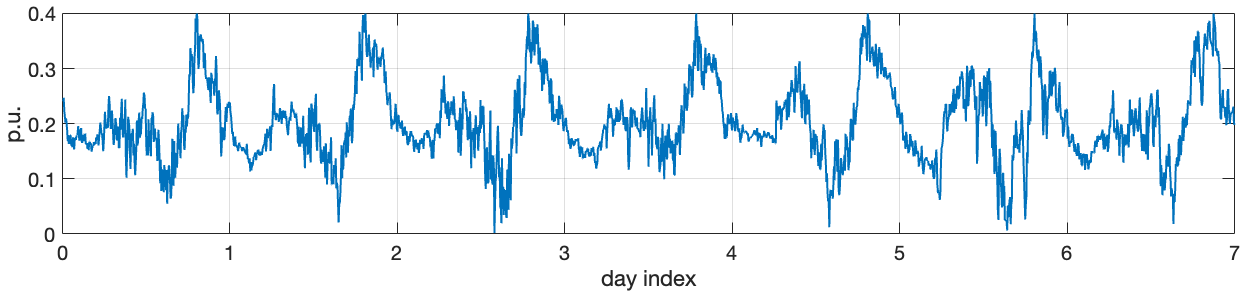}
\label{fig:Load}}
\caption{Normalized PV generation and load profiles for seven days. This profile is used to generate PV and load generation for each node by multiplying with the nominal capacities.}
\label{fig:PV_load}
\vspace{-1em}
\end{figure}


\begin{figure}[!htbp]
\centering
\subfloat[Unfair case.]{\includegraphics[width=0.5\linewidth]{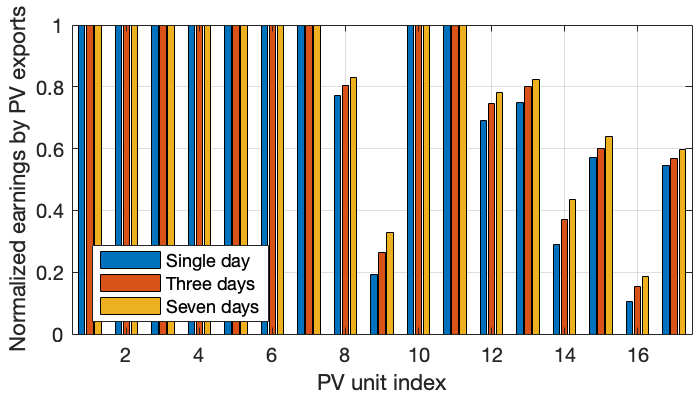}
\label{fig:1day}}
\subfloat[Fair case.]{\includegraphics[width=0.5\linewidth]{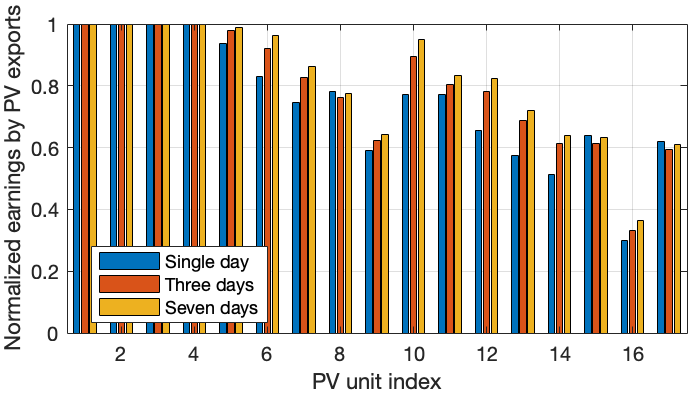}
\label{fig:3day}}
\caption{Normalized earnings by exported PV generation with different days of simulations for (a) unfair case and (b) fair case by feedback-based scheme.}
\vspace{-0.5em}
\label{fig:bill_results}
\end{figure}

\begin{figure}[!htbp]
\centering
\subfloat[Unfair case.]{\includegraphics[width=0.5\linewidth]{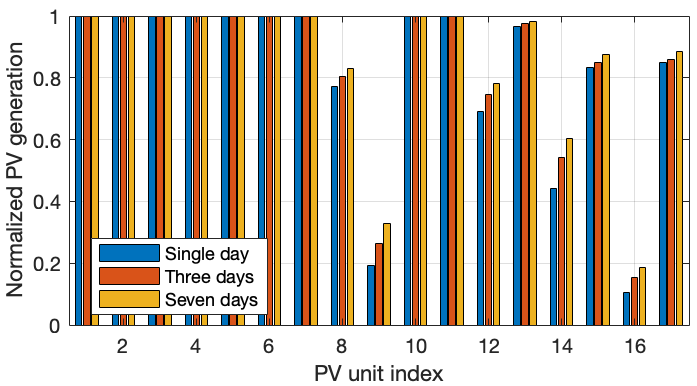}
\label{fig:1day_curt}}
\subfloat[Fair case.]{\includegraphics[width=0.5\linewidth]{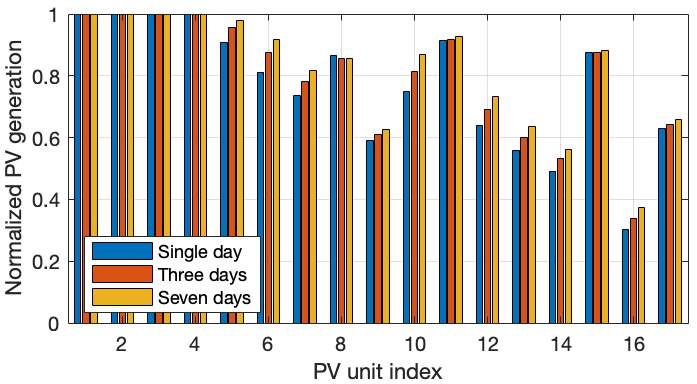}
\label{fig:3day_curt}}
\caption{Normalized PV generation with different days of simulations for (a) unfair case and (b) fair case by feedback-based scheme.}
\label{fig:curtail_results}
\vspace{-0.5em}
\end{figure}
\subsubsection{Other Testcases} We also simulate the feedback-based fairness scheme on larger distribution systems, specifically case33~\cite{baran1989network}, case69~\cite{baran1989network}, and case141~\cite{khodr2008maximum}. 

To simulate the proposed algorithm, we require time-series data of the PV generation and demand. We model these using historical measurements from the microgrid experimental setup at the EPFL's Distributed Electrical Systems Laboratory detailed in \cite{gupta2020grid}. We simulate for an extreme case, i.e., a low-demand scenario with high PV generation. Fig.~\ref{fig:PV_load} shows the profile of the PV generation and the load realizations. These profiles are shown for seven days. To generate the load and PV profiles per location in the network, we multiply with the nominal loads from the corresponding testcase. We use $c^\text{im}$ and $c^\text{fit}$ as 0.3 and 0.1 USD/kWh, respectively, for the numerical simulation based on a survey of US electricity prices.

%

\subsection{Fairness Metrics}
We evaluate fairness using two different metrics. The first is the Jain Fairness Index (JFI) \cite{jain1984quantitative} which quantifies the spread of benefits to each consumer with values ranging between 0 and 1, where JFI = 0 and JFI = 1 refer to completely unfair and fair cases, respectively. The JFI is:
\begin{align}
    \text{JFI}(x_1, \dots, x_l) = \frac{(\sum_{l\in\mathcal{N}_\text{pv}}x_l)^2}{|\mathcal{N}^{\text{pv}}|{\sum_{l\in\mathcal{N}_\text{pv}}x_l^2}}.
\end{align}

The second is called the Gini index, which is widely used for quantifying income inequality. The index is defined as:
\begin{align}
    \text{Gini}(x_1, \dots, x_l)  = \frac{\sum_{l\in\mathcal{N}_\text{pv}}\sum_{m\in\mathcal{N}_\text{pv}}|x_l - x_m|}{2|\mathcal{N}^{\text{pv}}|\sum_{l\in\mathcal{N}_\text{pv}}x_l}.
\end{align}
The Gini coefficient's value ranges from 0 (total fairness) to 1 (total inequality). 

The JFI and Gini metrics are computed for the two discussed objectives using $\mathcal{E}_l$ and $\mathcal{G}_l$ in \eqref{eq:fairness_bill_func} and \eqref{eq:fairness_curt_func}, respectively.
\subsection{Results with Feedback Controller}
\subsubsection{Cigre LV}
The results are shown in Figs.~\ref{fig:bill_results} and \ref{fig:curtail_results} for the objective of minimizing electricity bills and curtailments, respectively. We show the results for different days of simulations to see the evolution of fairness over the days. The results are shown for the simulation of one, three, and seven days in three different color bars. Figs.~\ref{fig:bill_results}-\ref{fig:curtail_results}~(a), and Figs.~\ref{fig:bill_results}-\ref{fig:curtail_results}~(b) show results for the ``unfair'' and ``fair'' cases, respectively. Observe that the schemes with fairness-informed control attain a fairer distribution of the exports earning and PV generation across the PV plants compared to the ``unfair'' case. In particular, PV units 9, 14, and 16 increase their benefits and generation in the fair case to be closer to the other PV units.
Also, note that the disparities in the PV units' production and benefits decrease as we simulate longer periods. 

To quantitatively compare the proposed method, we also report the results of net curtailments, ``Curtail'' in percentage, and fairness metrics in Tables~\ref{tab:bill} and \ref{tab:curt}, respectively. Again, we show the results for different numbers of consecutive days of simulations. Fairness is reported in JFI and Gini defined earlier. Observe that the JFI and Gini values improves (indicate increasing fairness via increases in JFI and decreases in Gini) for longer simulations. These simulations demonstrate how the proposed method improves the fairness of the PV curtailment schemes. 

\begin{table}[t]
    \centering
    \caption{Performance evaluation of the \emph{feedback-based approach for fairness} and comparison against the unfair case for the \emph{electricity bill objective}.}
    \begin{tabular}{|c|c|c|c|c|c|c|}
    \hline
    & \multicolumn{3}{|c|}{\bf Unfair } & \multicolumn{3}{|c|}{\bf Fair }\\
    \hline
    \bf{Days}  & \bf Curtail  & \bf JFI & \bf Gini & \bf Curtail  & \bf JFI & \bf Gini \\
    \hline
          1 &  29.4~\% & 0.86 & 0.21 & 31.8~\% & \bf 0.94 & \bf 0.14\\
          \hline
          3 & 26.4~\% & 0.89 & 0.18 & 27.8 ~\%& \bf 0.95 & \bf 0.13\\
          \hline
          7 & 24.2~\% & 0.88 & 0.17 & 25.3~\% & \bf 0.95 & \bf 0.12\\
        \hline
    \end{tabular}
    \label{tab:bill}
\end{table}
\begin{table}[t]
    \centering
    \caption{Performance evaluation of the \emph{feedback-based approach for fairness} and comparison against the unfair case for the \emph{PV curtailment objective}.}
        \begin{tabular}{|c|c|c|c|c|c|c|}
    \hline
    & \multicolumn{3}{|c|}{\bf Unfair } & \multicolumn{3}{|c|}{\bf Fair }\\
    \hline
    \bf{Days}  & \bf Curtail  & \bf JFI & \bf Gini & \bf Curtail  & \bf JFI & \bf Gini \\
    \hline
          1 & 29.4~\% & 0.81 & 0.25 & 32.0~\% & \bf 0.91 & \bf 0.18 \\
          \hline
          3 & 26.4~\% & 0.84 & 0.22 & 28.4~\% & \bf 0.93 & \bf 0.16 \\
          \hline
          7 & 24.2~\% & 0.87 & 0.20 & 25.8~\% & \bf 0.93 & \bf 0.15\\
        \hline
    \end{tabular}
    \label{tab:curt}
\end{table}


\subsubsection{Other Testcases} The results are shown in Tables~\ref{tab:testcase_bill} and \ref{tab:testcase_curt} for two different objectives. These tables show that the feedback-based scheme performs better in terms of fairness compared to the usual unfair case for all these test networks. 
\begin{table}[!t]
    \centering
    \caption{Performance evaluation on \emph{different test-cases} of the \emph{feedback-based approach for fairness} and comparison against the unfair case for the \emph{electricity bill objective}.}
        \begin{tabular}{|c|c|c|c|c|c|c|}
    \hline
    & \multicolumn{3}{|c|}{\bf Unfair } & \multicolumn{3}{|c|}{\bf Fair }\\
    \hline
    \bf{Testcases}  & \bf Curtail  & \bf JFI & \bf Gini & \bf Curtail & \bf JFI & \bf Gini \\
    \hline
          case33 & 32.6~\% & 0.89 & 0.18  & 34.9~\%  & \bf 0.90  & \bf 0.18 \\
          \hline
          case69 & 31.0~\% & 0.91 & 0.15  &  33.4~\% & \bf 0.94 & \bf 0.14 \\
          \hline
          case141 &  27.6~\% & 0.92 & 0.15 & 28.6~\% & \bf 0.96 & \bf 0.12 \\
        \hline
    \end{tabular}
    \label{tab:testcase_bill}
\end{table}
\begin{table}[!t]
    \centering
    \caption{Performance evaluation on \emph{different test-cases} of the \emph{feedback-based approach for fairness} and comparison against the unfair case for the \emph{PV curtailment objective}.}
        \begin{tabular}{|c|c|c|c|c|c|c|}
    \hline
    & \multicolumn{3}{|c|}{\bf Unfair } & \multicolumn{3}{|c|}{\bf Fair }\\
    \hline
    \bf{Testcases}  & \bf Curtail & \bf JFI & \bf Gini & \bf Curtail & \bf JFI & \bf Gini \\
    \hline
     case33 & 32.6~\% & 0.80 & 0.25  & 35.6~\%  & \bf 0.85 & \bf 0.23 \\
          \hline
          case69 & 30.9~\% & 0.81 & 0.23  &  33.4~\% & \bf 0.90 & \bf 0.19 \\
          \hline
          case141 & 27.6~\% & 0.84 & 0.22  & 28.9~\% & \bf 0.91 & \bf 0.18\\
        \hline
    \end{tabular}
    \label{tab:testcase_curt}
\end{table} 

\subsection{Extra Fairness Objective with and without Feedback}
In this section, we compare the schemes with additional fairness objectives. 
Here, we consider four formulations:
\begin{enumerate}
    \item Additional fairness objective \emph{without feedback-based approach} (Sec.~\ref{sec:fairness_w_obj}) with $h_l$ \emph{not accounting for past curtailments}, i.e., defined as \eqref{eq:objective_w_fairnessp0}. The case is referred to as \textbf{F0P0}. 
    \item Additional fairness objective \emph{without feedback-based approach} (Sec.~\ref{sec:fairness_w_obj}) with $h_l$ \emph{accounting for past curtailments}, i.e., defined as \eqref{eq:objective_w_fairness}. The case is referred to as \textbf{F0P1}.  
    \item Additional fairness objective \emph{with feedback-based approach} (Sec.~\ref{sec:combined_formulation}) with $h_l$ \emph{not accounting for past curtailments}, i.e., defined as \eqref{eq:objective_w_fairness_curtp0}. The case is referred to as \textbf{F1P0}. 
    \item Additional fairness objective \emph{with feedback-based approach} (Sec.~\ref{sec:combined_formulation}) with $h_l$ \emph{accounting for past curtailments}, i.e., defined as \eqref{eq:objective_w_fairness_curtp1}. The case is referred to as \textbf{F1P1}. 
\end{enumerate}

We present a comparison for the CIGRE LV system for a single day of simulation. 
We show the results in the form of Pareto Curves that we obtain by varying the weight parameter $w$ in the fairness objective term.  

\begin{figure}[!htbp]
\centering
\subfloat[]{\includegraphics[width=0.5\linewidth]{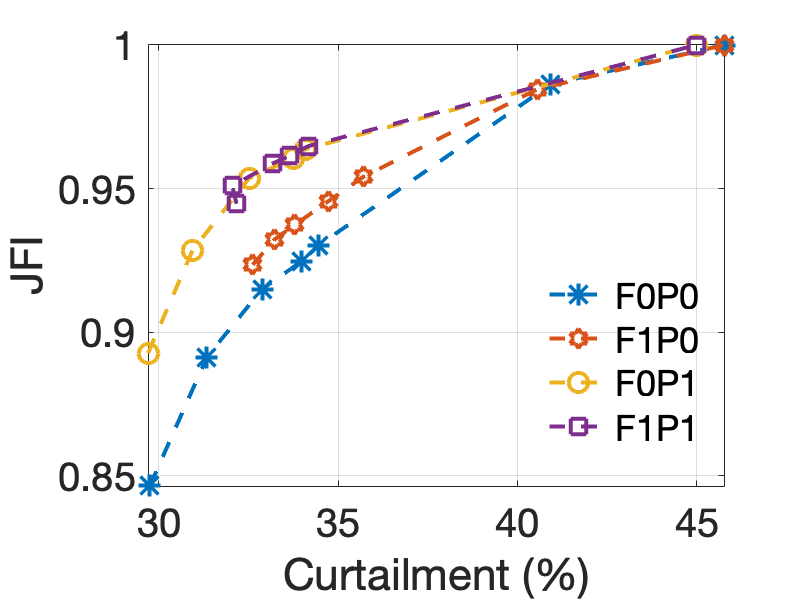}
\label{fig:Pareto_JFI_bill}}
\subfloat[]{\includegraphics[width=0.5\linewidth]{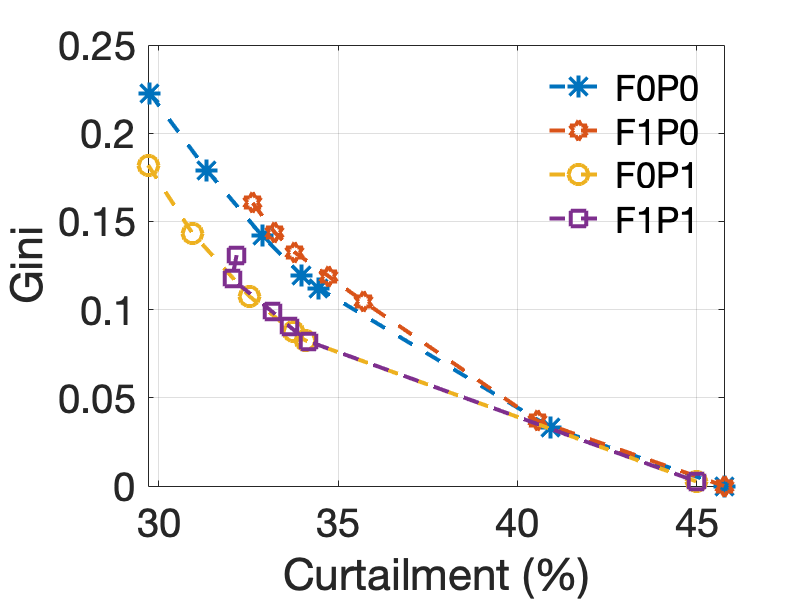}
\label{fig:Pareto_Gini_bill}}
\caption{Pareto curves depicting fairness (a) JFI and (b) Gini versus PV curtailments for the objective of minimizing electricity bills. Fairer and lower curtailment corresponds to the upper left for the JFI curve and the lower left for the Gini curve.}
\label{fig:Pareto_JFIGini_bill}
\end{figure}

\begin{figure}[!htbp]
\centering
\subfloat[]{\includegraphics[width=0.5\linewidth]{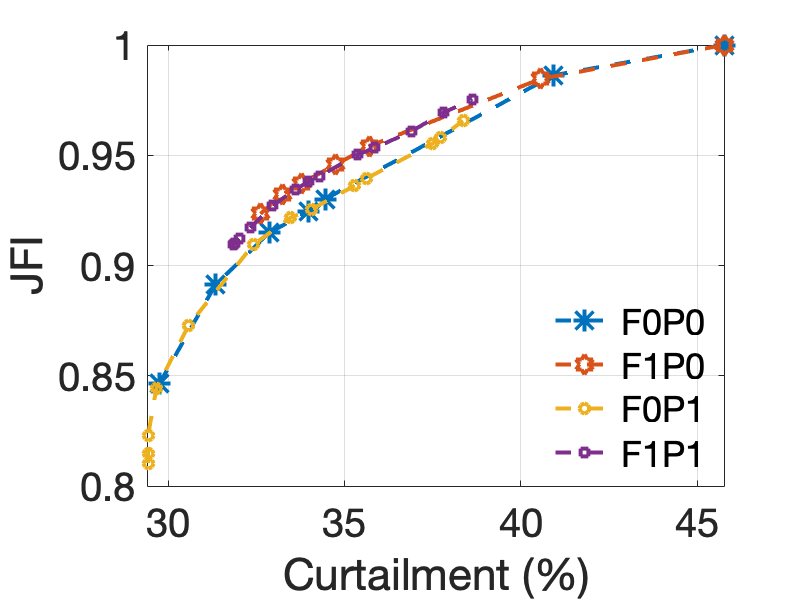}
\label{fig:Pareto_JFI_curt}}
\subfloat[]{\includegraphics[width=0.5\linewidth]{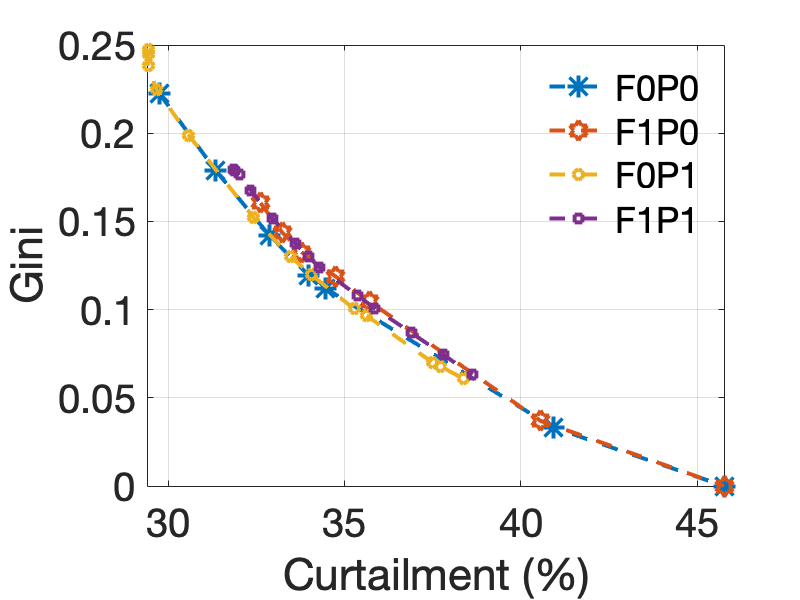}
\label{fig:Pareto_Gini_curt}}
\caption{Pareto curves depicting fairness (a) JFI and (b) Gini versus PV curtailments for the objective of minimizing PV curtailments. Fairer and lower curtailment corresponds to the upper left for the JFI curve and the lower left for the Gini curve.}
\label{fig:Pareto_JFIGini_curt}
\end{figure}

The results are shown for the two objectives in Figs.~\ref{fig:Pareto_JFIGini_bill} and \ref{fig:Pareto_JFIGini_curt}, respectively. These figures have fairness (quantified by JFI and Gini metrics) on the vertical axis and the net PV curtailments on the horizontal axis for different values of the weights assigned to the fairness term. 

For the plots for the electricity bill objective (Fig.~\ref{fig:Pareto_JFIGini_bill}), we vary the weight $w$ as $w = [0.001, 0.002, 0.003, 0.004, 0.005, $ $ 0.1, 0.2, 0.3, 0.4, 0.5]$. Observe that the scheme with no feedback and no past information on the curtailments (F0P0) is the least fair for any given value of PV curtailment. With the addition of past curtailment actions (F0P1), it achieves the same level of fairness with less PV curtailments, as seen in the JFI curve for F0P1, which is always above the one corresponding to F0P0, and vice versa for the Gini values.

We also show results when feedback is included, i.e., F1P0 and F1P1. We observe that the one with feedback (F1P1) achieves better fairness (i.e., higher JFI and lower Gini).


We also present an analysis for the curtailment objective in Fig.~\ref{fig:Pareto_JFIGini_curt}. For this Pareto curve, we vary the weight $w$ as $w = [0.01, 0.05, 0.1, 1, 2, 3, 4, 5, 6, 7, 8, 9, 10]$. In this case, we found that the Pareto curves coincide with each other for the cases when past curtailment information is included or not included, i.e., curves F0P1 and F0P0 coincide with each other, similarly for F1P0 and F1P1. This suggests that adding past curtailment information does not have much influence on the fairness values. This is contrary to what we observed in the case of electricity bill objective.

The JFI curve with feedback (F1P0 or F1P1) is always above the one without feedback (F0P0 or F0P1). This means that we attain a similar level fairness for the lower amount of PV curtailments when feedback is included. However, this outcome does not hold true if we compare the Gini metric, for which the Gini values suggest less fair (higher Gini) for the case with feedback compared to the one without. This suggests that it might be important to determine which other metrics can be used to evaluate fairness (see, e.g., \cite{Lan_axiomatic2010}) concerning different fairness objectives. Future work will investigate this seeming contradiction between the JFI and Gini fairness metrics.

\section{CONCLUSIONS AND FUTURE WORK}
\label{sec:conclusion}
This work compared two different approaches for increasing fairness in photovoltaic curtailments resulting from voltage control problems in power distribution grids. The first approach corresponds to maximizing an additional fairness term in the objective weighted by a factor. The second corresponds to a feedback-based approach where fairness is accounted for by assigning different weight factors based on previous timestep curtailments. We combine these two methods and perform analysis by varying different fairness functions and weights assigned to them.

Analyses were presented for several benchmark networks. We first compared the feedback-based approach against the unfair scheme.
The results consistently showed that the proposed scheme improved fairness, as quantified by Jain and Gini fairness metrics. Additionally, it was demonstrated that fairness improved with an increasing duration of simulations. 

Then, we performed a comparison between the schemes with additional fairness objectives with and without feedback consideration. The results showed that JFI has a higher value, i.e., better fairness, for the same level of PV curtailments. However, it was also observed that Gini values increase with the feedback-based approach, suggesting it to be less fair. This could be due to differences in the definition of the Jain and Gini metrics, and therefore it might be important to determine which metrics are used to evaluate fairness. Future work will investigate this difference in the behaviour of JFI and Gini fairness metrics.

We also studied the influence of including past curtailment information in the fairness function. This analysis showed that adding past information also improves fairness for the electricity bill objective, but not significantly for the curtailment minimization objective. This could be due to differences between load and the PV generation capacities per node, leading to a dominant local effect on the earnings of each PV plant. 


\bibliographystyle{IEEEtran}
\bibliography{biblio.bib}

\end{document}